\documentclass{iau}
\usepackage{graphicx}
\usepackage{natbib}
\usepackage{textcomp}
\usepackage{epsfig}
\usepackage{amsmath}
\usepackage{hyperref}

\title[Thermal characteristics of SOL2009-07-04T04:37 (B8.3) flare]
{Thermal characteristics of a B8.3 flare observed on July 04, 2009}

\author[Awasthi, A. K.; Sylwester, B.; Sylwester, J. and Jain, R.]   
{Arun Kumar Awasthi$^1$,
  Barbara Sylwester$^2$,
  Janusz Sylwester$^2$
  \and Rajmal Jain$^3$}

\affiliation{$^1$Astronomical Institute, University of Wroclaw, Poland\\ email: {\tt arun.awasthi.87@gmail.com} \\[\affilskip]
$^2$Space Research Center of Polish Academy of Sciences, Wroclaw, Poland\\email: {\tt bs@cbk.pan.wroc.pl, js@cbk.pan.wroc.pl} \\[\affilskip]
$^3$Kadi Sarva Vishwavidyalaya, Gandhinagar, India\\email: {\tt rajmal\_9@yahoo.com}}

\pubyear{2015}
\volume{320}  
\jname{Solar and Stellar Flares and Their Effects on Planets}
\editors{A.G. Kosovichev, S.L. Hawley, \& P. Heinzel, eds.}

\begin{document}

\maketitle

\begin{abstract}
We explore the temporal evolution of flare plasma parameters including temperature ($T$) - differential emission measure ($DEM$) relationship by analyzing high spectral and temporal cadence of X-ray emission in 1.6-8.0 keV energy band, recorded by SphinX (Polish) and Solar X-ray Spectrometer (SOXS; Indian) instruments, during a B8.3 flare which occurred on July 04, 2009. SphinX records X-ray emission in 1.2-15.0 keV energy band with the temporal and spectral cadence as good as 6 $\mu$s and 0.4 keV, respectively. On the other hand, SOXS provides X-ray observations in 4-25 keV energy band with the temporal and spectral resolution of 3 s and 0.7 keV, respectively. We derive the thermal plasma parameters during impulsive phase of the flare employing well-established Withbroe-Sylwester DEM inversion algorithm.
\keywords{Sun: corona, Sun: flares, plasmas, Sun: X-rays, radiation mechanisms: thermal, techniques: spectroscopic.}
\end{abstract}

\firstsection
\section{Introduction}
Thermal characteristics of solar flare plasma employing the multi-wavelength observations is of immense interest as it can shed light on the ongoing coupling processes in solar atmosphere. In particular, X-ray emission during a flare is the best probe of various thermal and non-thermal energy release processes (\cite{Brown1971}). Generally, flare plasma parameters viz. temperature ($T$), emission measure ($EM$), etc. are derived by forward-fitting/inversion of the observed X-ray spectrum (\cite{Jain2011}). However, the spectroscopic inversion of X-ray emission is an ill-posed problem, leading to substantial uncertainties in the derived $T$ and $EM$ values (\cite{Craig1976}). Moreover, several different $DEM$ inversion techniques, with various functional dependence of $DEM$ on $T$ viz. power-law, single-gaussian etc., are used to interpret observed X-ray spectrum. Further, Withbroe-Sylwester (W-S) $DEM$ inversion algorithm (\cite{Sylwester1980}, \cite{Kepa2008}) provides a more general scheme for such studies.
\\
Therefore, in this paper, we present the analysis of X-ray emission observed during a B8.3 flare occurred on July 04, 2009, the only event recorded in common with Solar X-ray Spectrometer (SOXS; \cite{Jain2005}) and Solar Photometer in X-rays (SphinX; \cite{Gburek2013}). Section 2 presents the observations while data analysis and results are given in Section 3. Section 4 presents the summary and conclusions.

\section{Observations}
\begin{figure*}[!htbp]
\begin{tabular}{cc}
    \epsfig{file=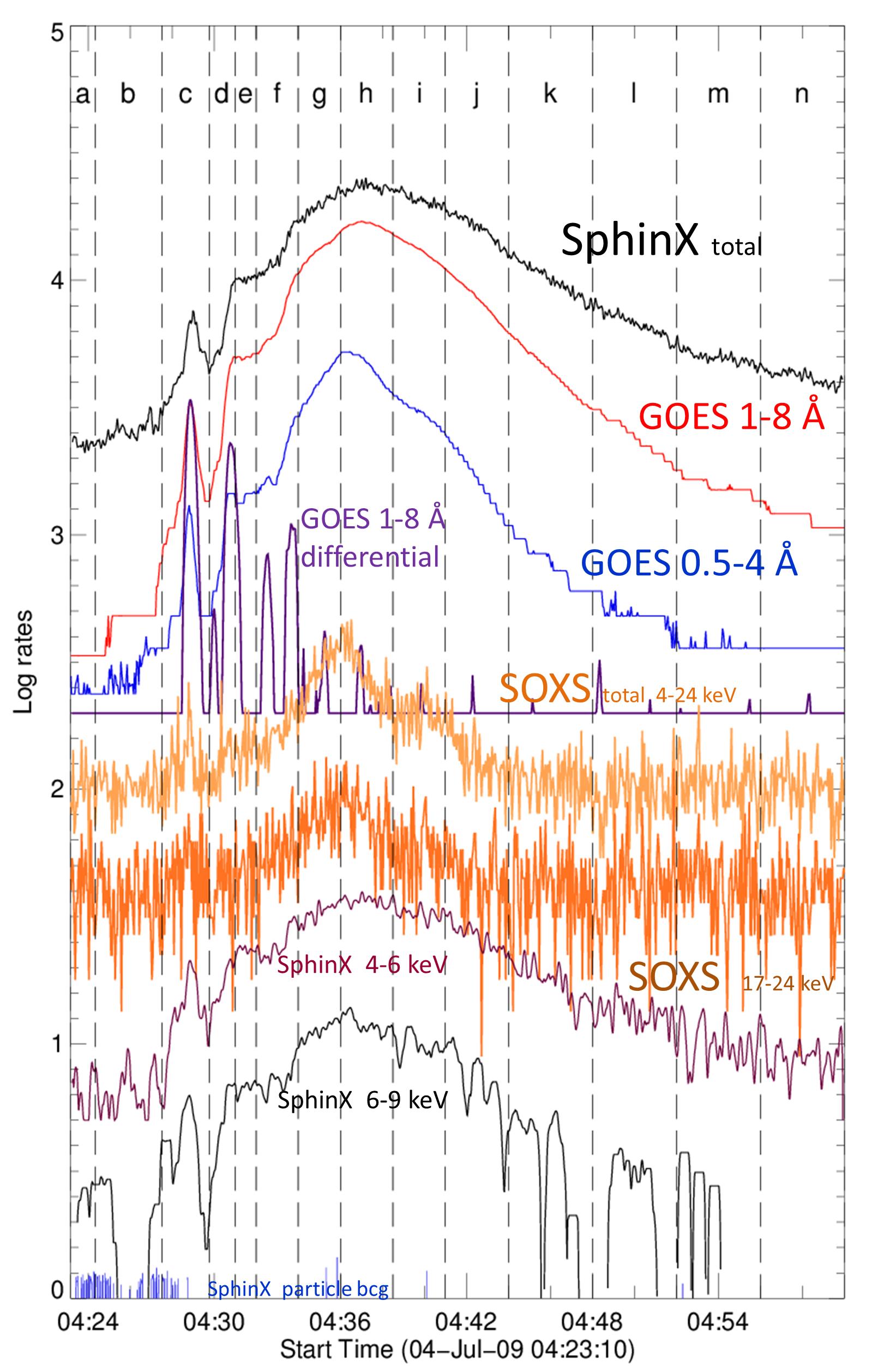, width=0.555\textwidth,  angle=0} &
    \epsfig{file=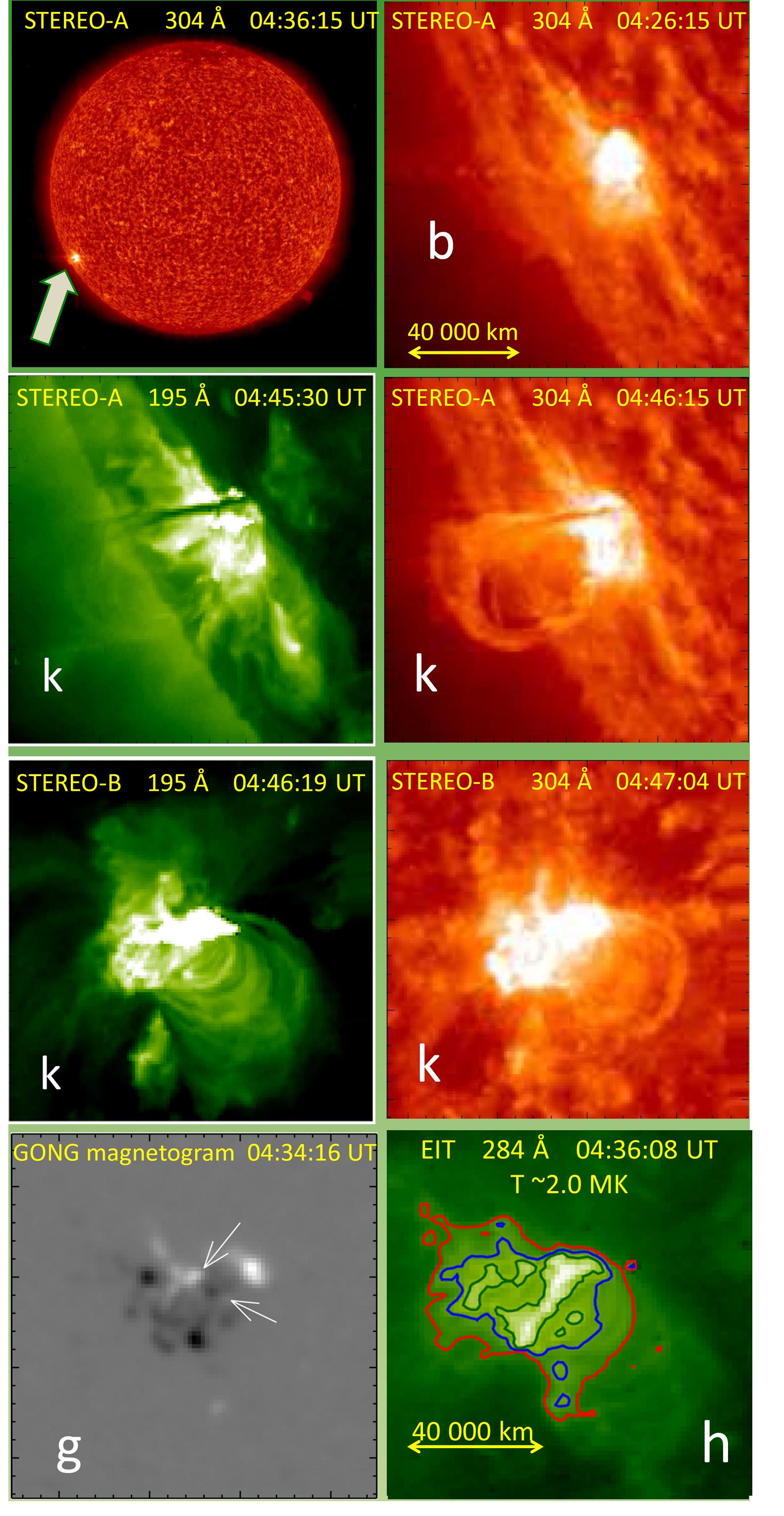, width=0.445\textwidth, angle=0}
\end{tabular}
\caption{Left Panel: Temporal evolution of X-ray emission as recorded by SphinX, SOXS and \textit{GOES} during SOL2009-07-04T04:37 (B8.3) flare. Dotted bars show the time intervals for which spectral analysis is undertaken. Right Panel: Multi-wavelength overview of the flare from \textit{STEREO-A} and \textit{STEREO-B} and EIT/\textit{SOHO}. Activity areas are shown by arrows in the GONG Magnetogram (bottom).}
\label{xr-lc}
\end{figure*}

We study a B8.3 flare event of July 04, 2009, which occurred in active region 11024. Thermal characteristics of the flare plasma are derived by analyzing X-ray spectra in 1.6-5.0 keV and 5.0-8.0 keV energy bands, recorded by SphinX and SOXS, respectively. Temporal evolution of X-ray emission during the flare as observed by SphinX and SOXS instruments as well as by \textit{GOES} is shown in the left panel of the Fig. \ref{xr-lc}. Further, morphological evolution of the flaring region is studied from the EUV images obtained from \textit{STEREO-A, B} and Extreme Ultraviolet Imager Telescope (EIT) onboard \textit{SOHO} mission, as shown in the right panel of the Fig. \ref{xr-lc}.

\section{Thermal characteristics of the flare plasma}
\begin{figure*}[!htbp]
 \begin{center}
    \includegraphics[width=0.95\textwidth, angle=0, clip]{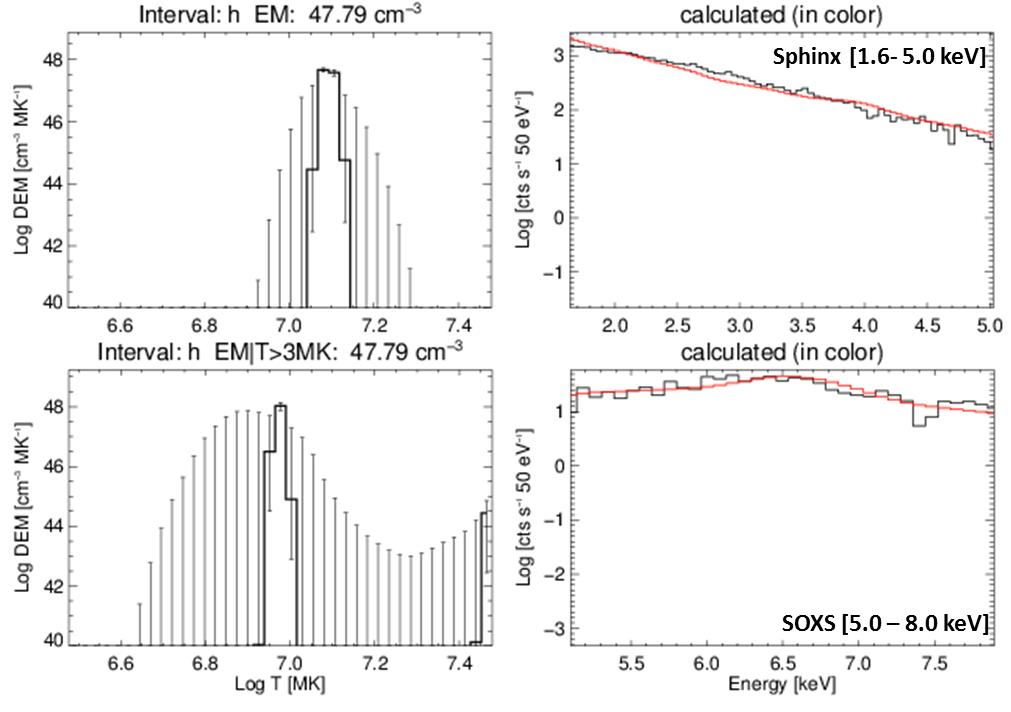}
    \caption{Top row: Best-fit \textit{DEM-T} relationship as well as the spectral-fit (drawn by red color) employing W-S inversion algorithm for the emission in 1.6-5.0 keV (plotted by black color), recorded by SphinX during 04:36:00-04:38:30 UT. Bottom row: Best-fit \textit{DEM-T} and fitted SOXS spectrum in 5.0 - 8.0 keV for the aforesaid time duration.}
    \label{sphinx-soxs-dem}
    \end{center}
 \end{figure*}

We analyze the X-ray spectra, recorded during the flare, with the help of W-S $DEM$ inversion method (\cite{Sylwester1980}). This numerical method employs maximum likelihood approach in which a $DEM-T$ distribution and hence corresponding theoretical spectrum is derived in an iterative manner with the aim to minimize its difference with the input observed spectrum (\cite{Kepa2008}). Coronal abundances are adopted from CHIANTI atomic database (\cite{Del Zanna2015}) while deriving the shape of theoretical spectra. As an input to this method, we have used fluxes recorded in the 73 energy bins (corresponding to the energy band 1.6-5.0 keV) and 35 energy bins (corresponding to the energy band 5.0-8.0 keV) by SphinX and SOXS instruments, respectively. The X-ray spectra are analyzed for various time duration as shown by dotted lines in the left panel of Fig. \ref{xr-lc}. Top row of the Fig. \ref{sphinx-soxs-dem} shows the best-fit $DEM-T$ relation derived by analyzing X-ray spectrum in 1.6-5.0 keV (low-energy), observed by SphinX during the peak of the impulsive phase of the flare, 04:36:00-04:38:30 UT. Similarly, in the bottom panel, we present the best-fit $DEM-T$ curve and spectral-fit over the X-ray spectrum in 5.0-8.0 keV (high-energy), observed by SOXS during the aforesaid time.

From Fig. \ref{sphinx-soxs-dem}, it may be noted that the best-fit $DEM-T$ relation derived from SphinX observation suggests nearly isothermal nature of the $DEM$, with the peak at temperature ($T_p$)$\sim$ 13 MK. Similarly, the best-fit $DEM$ to the SOXS spectrum in 5.0-8.0 keV energy band for the same time interval suggests isothermal nature in the form of single gaussian function dependence on $T$, however, at $T_p$ = 9.8 MK. Next, thermal energy are derived from the best-fit $DEM-T$ curve of Sphinx and SOXS observations and estimated to be 5.1 and 3.6 $\times$ $10^{29}$ ergs, respectively. We employ the volume estimated from EIT/\textit{SOHO} EUV wavelength images as shown in Fig. \ref{xr-lc} for the calculation of thermal energies.

\section{Summary and Conclusions}
We study the thermal characteristics of the plasma during SOL2009-07-04T04:37 (B8.3) flare by analyzing its X-ray spectrum in various energy bands, as obtained by SphinX and SOXS instruments. We summarize the preliminary findings of this study as follows:
\begin{enumerate}
  \item Emission-measure is found to be of isothermal nature during the peak of the impulsive phase of the flare.
  \item Thermal energy and the temperature estimated by analyzing low-energy (from SphinX) and high-energy (from SOXS) bands within SXR spectrum result in different peak temperature as well as thermal energy.
\end{enumerate}

In the next step, we have made a detailed investigation of thermal characteristics as well as the evolution of \textit{DEM-T} relationship in various phases of the flare by combining the observations from SphinX and SOXS instruments (\cite{Awasthi2016}). 

\textbf{Acknowledgements:} This research has been supported by Polish grant UMO-2011/01\\
/M/ST9/06096. Moreover, the research leading to these results has received funding from the European Community's Seventh Framework Programme (FP7/2007-2013) under grant agreement no. 606862 (F-CHROMA).

\end{document}